\documentclass[dvips,onecolumn,aps,showpacs,superscriptaddress,amssymb,pra,10pt]{revtex4}
\usepackage{graphicx}
\usepackage{dcolumn}
\usepackage{bm}
\usepackage{epsfig}
\usepackage{color}
\usepackage{longtable}
\usepackage{soul}
\usepackage{ulem}
\usepackage{amssymb,amsmath}
\usepackage{verbatim}
\usepackage{amsmath}
\usepackage{amssymb}
\everymath=\expandafter{\the\everymath\displaystyle}

\newcommand{\dd}{\mathrm{d}}

\renewcommand{\vec}[1]{\bm{#1}}

\newcommand{\abs}[1]{\left |{#1}\right |}
\def\etal{{\it et al.}}                                        
\begin{document}
\title{Analytical evaluation of atomic form factors: application to Rayleigh scattering} 
%
%
%
%
\author{L. Safari}
\email{laleh.safari@oulu.fi}
\affiliation{Department of Physics, University of Oulu, Box 3000, FI-90014 Oulu, Finland}

\author{J. P. Santos}
\affiliation{Centro de F\'{i}sica At\'{o}mica, Departamento de F\'{i}sica, Faculdade de Ci\^{e}ncias e Tecnologia, FCT, Universidade Nova de Lisboa, P-2829-516 Caparica, Portugal}

\author{P. Amaro}
\affiliation{Centro de F\'{i}sica At\'{o}mica, Departamento de F\'{i}sica, Faculdade de Ci\^{e}ncias e Tecnologia, FCT, Universidade Nova de Lisboa, P-2829-516 Caparica, Portugal}
\affiliation{Physikalisches Institut, Universit\"{a}t Heidelberg, D-69120 Heidelberg, Germany}

\author{K. J\"ank\"al\"a}
\affiliation{Department of Physics, University of Oulu, Box 3000, FI-90014 Oulu, Finland}

\author{F. Fratini}
\affiliation{Department of Physics, University of Oulu, Box 3000, FI-90014 Oulu, Finland}
\affiliation{Universidade Federal de Minas Gerais, Instituto de Ci\^encias Exatas, Departamento de F\'isica, 31270-901 Belo Horizonte, MG, Brasil}
\affiliation{Institut N\'eel-CNRS, BP 166, 25 rue des Martyrs, 38042 Grenoble Cedex 9, France}
%
%
%
%
\date{\today \\[0.3cm]}%
%
%
%
%
\begin{abstract}
Atomic form factors are widely used for the characterization of targets and specimens, from crystallography to biology. By using recent mathematical results, here we derive an analytical expression for the atomic form factor within the independent particle model constructed from nonrelativistic screened hydrogenic wavefunctions. The range of validity of this analytical expression is checked by comparing the analytically obtained form factors with the ones obtained within the Hartee-Fock method. As an example, we apply our analytical expression for the atomic form factor to evaluate the differential cross section for Rayleigh scattering off neutral atoms.
\end{abstract}
%
%
%
%
\maketitle
%
%
%
%
%
\section{Introduction}
Form factors (FFs) of specimens of various types, from nuclear compounds to biological tissues, have been widely investigated in the past and also recently in different branches of science, since they provide detailed information concerning the electric charge distribution of the specimens \cite{Seth2013, Ahmed2012, Bernauer2011, CoF2004, Chantler2000, Tar2002}. By systematically measuring FFs, information about the electric charge distribution of targets has been extensively extracted from scattering experiments \cite{BaC1961, AlS2008}. Atomic FFs, especially, play an important role in different topics of applied science such as: radiation absorption in shielding and medical diagnostics \cite{Ces1992}, structural factors of crystals in crystallography \cite{Giac, Wil2009}, and image contrast in Transmission Electron Microscopy \cite{Wil2009}. 

During the last decades, extensive amount of numerical calculations on nonrelativistic FFs (NFFs) and relativistic FFs (RFFs) for atoms and ions have been carried out, mostly by using Hartree-Fock (HF) wave-functions. Such numerical calculations can be found in several tables \cite{Crystallography:92, J.H.Hubbell:75, J.H.Hubbell:79}. 
Simultaneously, there has been many attempts toward analytical evaluations of atomic FFs. For instance, Belkic \etal{} reported several works on analytical atomic FFs including both the hydrogen-like and Clementi-Roetti wave functions for multi-electron atomic systems \cite{D.Belkic:83,D.Belkic:84,D.Belkic:89,D.Belkic:92}. In addition, partial analytical evaluations of atomic FF has been reported, for instance, in Ref. \cite{R.T.Brown:70} and also Eq. (4.4.5.5) in Ref. \cite{Crystallography:92}.

In contrast, here, we show that it is possible to obtain an analytical expression for the atomic FF, alternatively, by using recent mathematical results: We derive an analytical expression for the NFF based on the Independent Particle Model (IPM) constructed from screened hydrogenic wave functions. To derive such an expression, we make use of the Jacobi-Anger expansion that yields a multipole decomposition of the NFF, with a full separation of the angular and radial parts. We then derive an analytical form for the radial part by means of recently discovered solutions of integrals involving the product of Bessel functions and associated Laguerre polynomials \cite{R.S.Alassar:08}. 
Such analytical solutions of atomic FFs based on multipole decomposition, to the best of our knowledge, have not been presented in the literature yet.  

Even though approximative, 
IPM is widely used in many fields of science in order to grasp basic properties of the system under investigation (e.g., in atomic and nuclear physics) and is always presented in textbooks as a first approach to describe many-body systems. Due to this and to the wide applicability of FFs, we state that our analytical expression has a fundamental value. Besides, such an analytical expression might have also computational value, as remarked in \cite{Schaupp:1983}.
First, numerical calculations of FF require specific programs (e.g. Cowan \cite{R.D.Cowan:81}) and programming capabilities, whereas an analytical approach can be implemented by nearly anyone in any symbolic software (e.g. Mathematica). Second, numerical calculations might involve problems due to rapid oscillation of the integrand at high energies, due to high multipoles, and are approach dependent. Therefore, they need to be checked for each element and energy value. On the other hand, an analytical formula is general and valid for any element. Third, numerical values of the FF in the literature for specific momentum-transfer-grids are limited and dependent on the particular technique of interpolation, while the analytical FF function can generate FF values for any selected momentum transfer. In this regard, empirical formulae for the FF have been obtained and largely used in the literature (see, for instance, Refs. \cite{W.Muhammad:2013, Bethe:1952, Smend:1974}, and also International Table of Crystallography (ITC) \cite{Crystallography:92}, where the atomic FF is analytically expressed by means of four gaussians whose parameters are obtained from least square fitting of numerical values). An analytical formula that can provide easy access to reliable FF values for a wide linear momentum range can thus be very profitable for users, especially for experimentalists. 

The range of validity of our analytical expression is assessed by comparing the analytically obtained FFs with the ones obtained within the HF method. 
As a use case example, we apply our analytical expression for the atomic FF to Rayleigh scattering by neutral atoms 
\cite{P.P.Kane:86,Chantler2000}. The differential cross section (DCS) for Rayleigh scattering off some selected neutral atoms is here calculated in two ways: i) by using our analytical expression for the FF based on IPM, and ii) by using the FF obtained from single-configuration HF numerical calculations. Both results are compared with experiments and other numerical calculations presented in the literature. Finally, it is shown how our analytical formula for the evaluation of the atomic FF can be implemented in configuration interaction (CI) numerical algorithms, extended to a relativistic framework, and applied to structural crystallography. 

SI units are used throughout this article.

%
%

\section{Atomic model}
\label{sec:AM}
Within the Independent Particle Approximation (IPA), the total wave function of the atom, $\Psi(\vec{r}_1,...,\vec{r}_N)$, is considered as product of $N$ one-electron spin orbitals in the form of one or more Slater determinants. The one-electron spin orbitals can be hydrogenic or non-hydrogenic for a self-consistent potential generated by the electrons and nucleus. The HF method, which is also called the self-consistent field (SCF) method, is based on IPM. In hydrogenic IPA, which we refer to as IPM, the total wave function is expressed by means of a Slater determinant constructed from hydrogenic solutions of the Schr\"odinger equation (i.e., Coulombian wavefunctions) \cite{B.H.Bransden:83,F.Fratini:2011,Andrey:2010}. In IPM, only the electron kinetic energies and the electron-nucleus Coulomb interaction terms are retained in the hamiltonian. However, electron-electron interactions are partially taken into account by using an effective nuclear charge, for each electronic orbital. The effective nuclear charge thus corresponds to the nuclear charge as seen by a given electron due to the screening of the other electrons. In what follows, the effective nuclear charge of the $i$th orbital will be denoted by $Z_i$. In our analytical method, values of $Z_i$ are directly taken from tabulation of Clementi \etal~ \cite{E.Clementi:1963,E.Clementi:1967}, which are currently available in the website of WebElements \cite{webelement}. Clementi \etal~ computed the self-consistent-field (SCF) function for atoms with a minimal basis set of Slater-type orbitals. The orbital exponent of the atomic orbitals, $\xi$, are optimized as to ensure the energy minimum. With such analysis, they obtained simple and accurate rules for the electronic screening constant, $\sigma$. Their rules accounts also for the screening due to the outside electrons in comparison to the Slater's rule \cite{Slater:1930}. Finally they obtained effective nuclear charge with formula: $Z_{\textrm{eff}}=n \xi= n (Z-\sigma)$, where $Z$ is the atomic number. 
Here, we shall use such $Z_i$, since Coulombian orbitals are well approximated by Slater-type orbitals \cite{A.Costescu:2011}.

The electronic charge distribution of atoms can be written as
\begin{eqnarray}
\label{eq:ChargDistrib}
\rho(r)\,=\,\frac{N}{4\pi}\,\int \dd\Omega_r\int\dd\vec{r}_2...\dd\vec{r}_N\,\abs{\Psi(\vec{r}, 
\vec r_2...,\vec{r}_N)}^2~,
\end{eqnarray}
where $\dd\Omega_r$ is the differential solid angle related to the variable $\vec r$, $N$ is the number of electrons and $\Psi(\vec{r}_1,...,\vec{r}_N)$ is the $N$-body wave function of the atom. As evident from Eq. \eqref{eq:ChargDistrib}, we take the distribution $\rho(r)$ to be normalized to $N$ when integrated over $\dd\vec r$.
Within IPM, the electronic charge distribution can be expressed as
\begin{eqnarray}
\label{eq:Rho-IPM}
\rho_{\mbox{\tiny IPM}}(r)\,=\,\frac{1}{4\pi}\sum^{N}_{i=1}\abs{R_{n_i l_i}(r)}^2~,
\end{eqnarray}
where $R_{n_i l_i}(r)$ are hydrogenic radial wave functions that have the form \cite{B.H.Bransden:83}
\begin{eqnarray}
\label{eq:R-r}
R_{n_i l_i}(r)\,=\,M_{n_i l_i}\,\left(\mbox{\small{$\frac{2Z_ir}{n_ia_0}$}}\right)^{l_i}\,
                   \mbox{\Large$e$}^{-\frac{Z_ir}{n_ia_0}} 
                \;\mbox{\Large$L$}^{2l_i+1}_{n_i-l_i-1}
                \left(\mbox{\scriptsize{$\frac{2Z_ir}{n_ia_0}$}}\right).
\end{eqnarray}
Here, $n_i$ and $l_i$ are the principle and orbital angular momentum quantum numbers respectively, while $M_{n_i l_i}$=\mbox{\footnotesize${\frac{2}{n_i^2}\left(\frac{Z_i}{a_0}\right)^{\frac{3}{2}} \sqrt{\frac{(n_i-l_i-1)!}{(n_i+l_i)!}}}$} is the normalization coefficient, $a_0$ is the Bohr radius of hydrogen atom and $L^{\alpha}_{n}(x)$ is the associated Laguerre polynomial of degree $n$.

In writing Eq. \eqref{eq:Rho-IPM}, we have neglected cross term of the type $\sim R_{n_il_i}R_{n_jl_i}$ which arise due to the fact that radial Coulombic orbitals with different effective nuclear charge are not orthogonal to each other. 

%
%
\section{Atomic form factor}

\subsection{Multipole expansion of the form factor of spherically symmetric charge distributions}
For spherically symmetric charge distributions, the FF is given by \cite{P.P.Kane:86}
\begin{eqnarray}
\label{eq:FF}
F(q)\,=\,4\pi\int_0^{\infty}{dr\,r^2\,\rho(r)\,\frac{\sin(qr)}{qr}}~,
\end{eqnarray}
where $\rho(r)$ is the target charge distribution, $\hbar q=2\hbar\,k\sin(\theta/2)$ is the modulus of the momentum transferred in an elastic scattering, $\hbar$ is the reduced Planck constant, and $\theta$ is the scattering angle (the angle between the incoming and outgoing projectile directions). 
Moreover, $\hbar k$ is the modulus of the linear momentum of the scattering projectile. 
By rewriting $\sin(qr)$ appearing in Eq. \eqref{eq:FF} in exponential form and by making use of the Jacobi-Anger expansion, we can re-express the FF in the simple form 
\begin{eqnarray}
\label{FF-theta-k-Z}
     F(q)\equiv F(\theta,k)=\,
     \frac{4\pi}{k}\sum^{+\infty}_{\substack{L=1\\ \textrm{(odd)}}} \frac{\sin\left(L\frac{\theta}{2}\right)}
     {\sin\left(\frac{\theta}{2}\right)}\,f_L(k)~,
\end{eqnarray}
where the summation over $L$ is restricted to odd numbers, as explicitly denoted. The radial part of the FF ($f_L$), which will be hereinafter referred to as ``radial form factor'', is defined as
\begin{eqnarray}
\label{ff-r-Z}
f_L(k)=\int^{\infty}_0\dd r\,r\,\rho(r)\,J_L(2kr)~.
\end{eqnarray}
The simple form for the FF in Eq. \eqref{FF-theta-k-Z} allows for a few physical considerations. Firstly, being $J_1$ the Bessel function of lowest order included in the summation, for $kr\ll 1$ one may use the equivalence $J_L(2kr)\approx kr\,\delta_{L1}$ and therefore the radial FF would be $Nk/(4\pi)$. Consequently, the FF would simply equal $N$, as it is expected to be for a point-like target whose charge is $N$ times the electronic charge. This is physically plausible, since for $kr\ll 1$ the resolution of the probe (projectile) is not enough to resolve the target structure or, in other words, the target is ``seen'' as point-like by the probe. Secondly, we can interpret the summation in Eq. \eqref{FF-theta-k-Z} as a multipole expansion of the FF, where each multipole is specified by the integer $L$. Such a multiple expansion can be used with any model and allows to calculate the FF with the necessary accuracy by setting the maximum allowed multipole. The angular and radial parts of the FF in this expansion are fully separated. Whereas the angular part is expressed analytically, the radial part is in integral form. Below we shall see that, within IPM constructed from screened hydrogenic wave-functions, the radial part can also be expressed analytically.

\subsection{Analytical expression for the atomic Form Factor within IPM}

\begin{figure}[t!]
\begin{center}
\includegraphics[
scale=0.5]{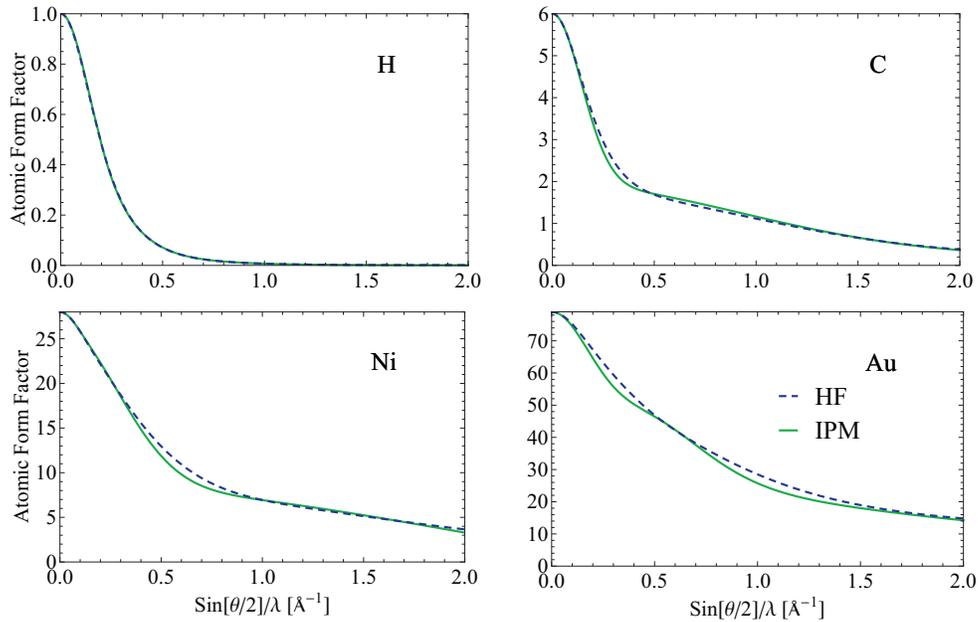}
\caption{(color online). Atomic form factors for selected elements. HF's and IPM's calculations are compared.
}
\label{FFfig}
\end{center}
\end{figure}

In this section, we shall proceed to evaluate the radial FF within the IPM. Such quantity shall be denoted by $f_L^{\mbox{\tiny IPM}}(k)$ and is given by Eq. \eqref{ff-r-Z} with the replacement $\rho(r)=\rho_{\mbox{\tiny IPM}}(r)$. Using Howell's formula (see Ref. \cite{W.T.Howell:37} or Eq.~(1) of Ref. \cite{L.Carlitz:61}), the square of the associated Laguerre polynomial in Eq. \eqref{eq:Rho-IPM} is decomposed into a sum of associated Laguerre polynomials. After such a decomposition, the function inside the integral \eqref{ff-r-Z} reduces to a product of a Bessel function and an associated Laguerre polynomial. Alassar \etal~recently derived the result for integrals involving such functions in terms of hypergeometric function (see Eq.~(6) of Ref.~\cite{R.S.Alassar:08}). By thus using Alassar's result and Eq.~\eqref{FF-theta-k-Z}, we find the following analytical expression for the FF:
%
%
\begin{eqnarray}
\label{eq:FF-thetakZ-final}
     F_{IPM}(\theta,k)\, &=&
     \frac{1}{k}\,\sum^{+\infty}_{\substack{L=1\\ \textrm{(odd)}}}\,
     \sum^{N}_{i=1}\:\sum^{n_i-l_i-1}_{\nu=0}\:\sum^{2\nu}_{m=0}\,
     \frac{\sin\left(L\frac{\theta}{2}\right)}{\sin\left(\frac{\theta}{2}\right)}\,
     \left(\frac{n_ia_0}{2Z_i}\right)^2\,\left(\frac{kn_ia_0}{Z_i}\right)^L\,\big(M_{n_il_i}\big)^{2}\,A^{\nu}_{n_il_i}
     \,B^{\nu mL}_{l_i}\, 
     \\ \nonumber
    && \times\; _2{\mbox{\large$F$}}_1\left(\,
    \frac{L+4l_i+m+3}{2}
        ,\,
    \frac{L+4l_i+m+4}{2}
        ;\,1+L;\,-\left(\frac{kn_ia_0}{Z_i}\right)^2\right)~,
\end{eqnarray}
%
%
where $_2{F}_1(a,\,b;\,c;\,z)$ is the Gaussian hypergeometric function. Here $A^{\nu}_{n_il_i}$ and $B^{\nu mL}_{l_i}$ are coefficients that can be directly traced back to Howell's and Alassar's formulas \cite{W.T.Howell:37,R.S.Alassar:08}, and are given by
\begin{eqnarray}
\label{eq:A}
A^{\nu}_{n_il_i}=\frac{(n_i+l_i)!\,(2\nu)!\,(2n_i-2l_i-2-2\nu)!}{2^{2n_i-2l_i-2}\,(n_i-l_i-1)!\,\nu !\,((n_i-l_i-1-\nu)!)^2\,(2l_i+1+\nu)!} 
\end{eqnarray}
and
\begin{eqnarray}
\label{B}
B^{\nu mL}_{l_i}= \frac{(-1)^m\,2^{m-L}\,(2\nu+4l_i+2)!\,(L+4l_i+m+2)!}{m!\,(2\nu-m)!\,(4l_i+m+2)!\,L!}
\end{eqnarray}
The Gaussian hypergeometric function in Eq. \eqref{eq:FF-thetakZ-final} can be evaluated analytically (see Eqs. (15.3.19) and (15.4.10) in Ref. \cite{Abramowitz}). Equation \eqref{eq:FF-thetakZ-final} is the main result of this article: By setting the orbital effective nuclear charges, the electronic configuration, the projectile wave-vector and the scattering angle, the atomic FF is analytically obtained. For example, in Fig. \ref{FFfig}, we show the FFs of some selected elements as obtained from \eqref{eq:FF-thetakZ-final}. The FFs are plotted against $\sin(\theta/2)/\lambda$, as usual in crystallography, where $\lambda=2\pi/k$. IPM's FFs are compared with HF's. For light atoms, FFs obtained within IPM are hardly distinguishable from the ones obtained within HF. 
As the atomic number increases, the difference between the two calculations is, as expected, more significant. This is due to the fact that the cross terms of the type $\sim R_{n_il_i}R_{n_jl_i}$, which have been here neglected in the calculation of the radial charge density \eqref{eq:ChargDistrib}, have non-negligible contribution in medium- and high-$Z$ regime when compared to the direct terms. 

As an example, below we shall apply Eq. \eqref{eq:FF-thetakZ-final} to evaluate the DCS for Rayleigh scattering by some selected neutral atoms. 

\begin{table*}
\begin{center}
\caption{Calculations for the DCS (barn/sr) for Rayleigh scattering by selected elements $1\leq Z\leq 82$ in the ground state, for scattering angle $\theta=90^\circ$, for selected energies. Our ANFF and HF results are compared with SM calculations \cite{P.P.Kane:86}, NNFF calculations \cite{J.H.Hubbell:75} and measurements. If not differently specified, the measurements of Mandal \etal~\cite{A.C.Mandal:02} are shown for photon energy $E_\gamma$ = 22.1 keV, while the measurements of Kumar \etal~\cite{S.Kumar:09} are shown for $E_\gamma$ = 59.54 keV. Numbers in parentheses are the experimental errors, while numbers in squared brackets are powers of ten.} \vspace{0.2 cm}
\label{Tab:I}
\begin{tabular}{lrrrrrrrrrrr}
\hline\hline \\[-0.1cm]
&\multicolumn{5}{c}{$E_\gamma$ = 22.1 keV}&&
\multicolumn{5}{c}{$E_\gamma$ = 59.54 keV} \\
\cline{2-6} 
\cline{8-12} \\[-0.2cm]
& & & &\multicolumn{2}{c}{this work}& & & & &\multicolumn{2}{c}{this work}\\
\cline{5-6}\cline{11-12}\\[-0.2cm]
$$Element$$&$$Measured$$&$$SM$$&$$NNFF$$&$$HF$$&$$ANFF$$&~~&$$Measured$$&$$SM$$&$$NNFF$$&$$HF$$&$$ANFF$$\\\hline \\[-0.1cm] 
$^{1}\textrm{H}$  &-           &-     &3.432[-7]&3.339[-7]&3.344[-7]&~~& -                   & -    &1.460[-10]&1.482[-7]&1.457[-10]\\
$^{13}\textrm{Al}$&  -         &-     &    0.133&0.120    &    0.115&~~&$0.01633^a$ (0.00035)&0.0165&    0.0169&0.0166   &0.01621   \\
$^{28}\textrm{Ni}$& 1.53 (0.13)& 1.560&    1.365&1.387    &    1.519&~~&      -              &-     &     0.106&0.097    &0.095     \\
$^{30}\textrm{Zn}$& 1.69 (0.15)& 1.745&    1.517&1.543    &    1.648&~~&       0.127 (0.006) &0.132 &     0.124&0.121    &0.112     \\
$^{46}\textrm{Pd}$& 4.20 (0.37)& 3.845&    5.206&5.364    &    4.188&~~&$0.555^b$ (0.028)    & 0.665&     0.562&0.591    &0.480     \\
$^{48}\textrm{Cd}$& 4.90 (0.44)& 5.019&    6.045&6.211    &    5.005&~~&        0.709 (0.035)& 0.752&     0.622&0.665    &0.553     \\
$^{50}\textrm{Sn}$& 5.82 (0.50)& 6.045&    6.879&7.087    &    6.101&~~&        0.759 (0.038)& 0.824&     0.697&0.734    & 0.632    \\
$^{79}\textrm{Au}$&18.40 (1.66)&20.468&   19.054&20.280   &   16.922&~~&         1.92  (0.10)& 1.93 &      2.27&2.560    & 1.90     \\
$^{82}\textrm{Pb}$&20.79 (1.83)&22.853&   21.683&23.055   &   18.509&~~&         2.38  (0.12)& 2.27 &      2.64&3.009    & 2.13    \\ \hline\hline
\multicolumn{9}{l}{$\substack{^{a}\textrm{Value taken from \cite{E.Casnati:90}.}}$}                                             \\
\multicolumn{9}{l}{$\substack{^{b}\textrm{Value taken from \cite{P.Latha:12}.}}$}                                               \\
\end{tabular}\end{center}
\end{table*}%
%
%
%
%
%
\section{Application to Rayleigh scattering}
\label{sec:Ray}
During the last decades, calculations of Rayleigh scattering amplitudes have been widely performed, both within relativistic and nonrelativistic frameworks \cite{L.Safari:12A,L.Safari:12B,L.Kissel:00,Royetal,P.P.Kane:86,L.Kissel:80,V.Florescu:90}. Up to now, there have been two main methods to perform such calculations, namely relativistic second-order S-matrix (SM) and FF approaches \cite{P.P.Kane:86}. 
SM approach is more accurate than FF, and can provide Rayleigh amplitudes as accurate as 1\%, for a wide range of photons energies (approximately from 50 keV to 1.4 MeV \cite{Roy1999}). Nevertheless, since precise SM calculations are computationally very demanding, FF approaches are very much used. A comparison between FF and SM approaches can be found elsewhere and will not be made here in detail (for details, see Refs. \cite{Royetal, Roy1999}).
We briefly mention that, although both SM and FF approaches have their own limitations depending on the energy of the incident photon, atomic number, atomic shell, and scattering angle, they provide DCS in good agreement with experimental data over limited intervals of these parameters \cite{Royetal}. More specifically, FF approach can be used provided that the photon energy be much larger than the atomic binding energies but much smaller than the electron rest-mass energy \cite{P.P.Kane:86}.
More accurate Rayleigh amplitudes for a wider photon energy range can be achieved within the FF approach by considering a modified form factor (MFF) and by considering the anomalous scattering factors usually denoted by $f'$ and $f''$ \cite{RoyKissel:1999, Chantler2000}. 

It must be underlined that both SM and FF approaches treat the target as an isolated atom (isolated atom approximation), with the electronic wave-functions being obtained within IPA. This restricts their use in situations where electron correlations are important, such as at very low photon energy and when the atom is influenced by the environment, as it might be the case in solids or in a plasma. For coherent scattering by several atoms in a unit cell, the more general Bragg-Laue (BL) or Thermal Diffuse (TD) diffraction laws must be used, which correspond to the cases where the Bragg planes are aligned or misaligned, respectively. Both BL and TD diffraction laws depend on the atomic form factor through the crystal structure factor \cite{Chantler2000}.

Within the FF approach (or FF approximation), the DCS for elastic scattering of photons by an isolated atom is given by \cite{P.P.Kane:86}
\begin{equation}
\label{eq:dsig_FF}
\frac{\dd\sigma}{\dd\Omega}\,=\,\frac{r_0^2}{2}\,(1\,+\,\cos^2\theta)\,\abs{F(\theta,k)}^2~,          
\end{equation}
where $\frac{r_0^2}{2}\,(1\,+\,\cos^2\theta)$ is the Thomson DCS, and $r_0$ is the classical electron radius. Below, we shall obtain the DCS for Rayleigh scattering by substituting \eqref{eq:FF-thetakZ-final} into \eqref{eq:dsig_FF}. In order to assess the validity of our analytical approach, we shall also numerically calculate the DCS by evaluating the FF directly from Eqs.~\eqref{eq:ChargDistrib}, \eqref{FF-theta-k-Z} and \eqref{ff-r-Z}, using wave functions obtained from single-configuration HF calculations, as implemented in Cowan's code \cite{R.D.Cowan:81}. We checked that HF and relativistic HF (so called {\it Dirac-Hartree-Fock} (DHF)) calculations give the same results for the FF, within few percent. We also checked that the influence of the Breit-Interaction in the calculation of the FFs is negligible.

We calculated the DCS for Rayleigh scattering off some selected elements in the ground state, for scattering angle $\theta=90^\circ$, for energies 22.1~keV and 59.54~keV. Our analytical nonrelativistic DCSs (hereinafter referred to as ANFF) and HF results are shown in Table \ref{Tab:I} and compared with SM calculations of Kissel \cite{P.P.Kane:86}, numerical NFF calculations of Hubbell \cite{J.H.Hubbell:75} (hereinafter referred to as NNFF), and recent measurements of Mandal~\etal~ \cite{A.C.Mandal:02} and Kumar~\etal~ \cite{S.Kumar:09}. 
As can be seen from Table \ref{Tab:I}, for the energy value 22.1~keV, ANFF cross sections are inside the experimental error bars for all elements except for $^{82}\textrm{Pb}$. 
In comparison, all SM cross sections are inside the error bars except for $^{79}\textrm{Au}$ and $^{82}\textrm{Pb}$ (for $^{46}\textrm{Pd}$, the cross section is in the border). On the other hand, NNFF and HF calculations are, in general, outside the experimental error bars and are in excellent agreement with each other. 
Another point which is worth to notice is that in the case of $^{46}\textrm{Pd}$, our ANFF cross section describes the measurement remarkably better than all other calculations. Comparison between ANFF and HF values shows that ANFF results agree better with experiments than HF results for photon energy $E_\gamma=22.1$ keV. The reason for this might be: i) In HF calculations, all parameters are calculated self-consistently. On the other hand, in the IPM the effective nuclear charge is given as external parameter, which is calculated independently. This difference in procedure might be the reason for which our analytical formula is favorable with respect to HF; ii) We must remember that the application of the FF to the DCS for light scattering is itself an approximation. Thus, the accuracy of FF, via either analytical or numerical methods, is not directly responsible for the discrepancies between experimental data and theoretical value.
For the energy value 59.54~keV, SM calculations have in general better agreement with experiments than other calculations. This is not unexpected, since SM calculations numerically include the full contribution of the Rayleigh scattering second order amplitude, while NNFF, HF and ANFF cross sections include only the leading part of it \cite{P.P.Kane:86}. 

\section{Further applications and prospects}
\label{sec:AppPro}
We would like here to outline few further theoretical analysis and applications which could be performed using the presented results: 

i) Our nonrelativistic analytical treatment can be extended to the calculation of RFF by using Dirac hydrogenic wave functions. In fact, radial components of Dirac hydrogenic wavefunctions can be expressed in terms of associated Laguerre polynomials. By repeating the same analysis showed in this article for each component, one can derive the relativistic equivalent of Eq. \eqref{eq:FF-thetakZ-final}. Relativistic effects in Rayleigh scattering off atoms are nonetheless expected to be small, as briefly discussed in Sec. \ref{sec:Ray} (see also Ref. \cite{RoyKissel:1999}, where relativistic effects in FF calculations are discussed to be not more than few percent, when momentum transfer is not very large).

ii) The presented formalism may be directly generalized to CI numerical calculations. The CI atomic state function (ASF) is a linear combination CSFs which, for instance, might be taken as screened hydrogenic wave functions. Accordingly, the FF related to the ASF is simply given by (neglecting cross terms)
\begin{equation}
F_{\mbox{\tiny CI}}(\theta,k)=\sum_i\,\abs{c_i}^2\,F_i(\theta,k)~,
\end{equation}
where $F_i(\theta,k)$ are the FFs presented in Eq. \eqref{eq:FF-thetakZ-final} for the $i$th CSF and $c_i$ are determined numerically by diagonalizing the Hamiltonian matrix. 

iii) In structural crystallography, the structure factor $F_{\vec h}$, which may be written as a linear combination of FFs of the atoms in the unit cell \cite{Giac, Wil2009}, is used to solve the crystallographic phase problem, and, therefore, to determine the crystal structure. Today the number of solved structures (inorganic, organic, metallorganic, biological tissues) is approaching one million and they are deposited in various databases \cite{CDS, ICDS, PDB}. The accuracy of the structural model is estimated by the crystallographic index, $R_{\textrm{cryst}}$, and then verified by means of chemical validation. Presently the most popular atomic FFs used in crystallography are listed in the International Tables for Crystallography \cite{Crystallography:92}. The analytical expression \eqref{eq:FF-thetakZ-final} may be used as a possible alternative.

iv) Our analytical expression for the atomic FF can be applied for studying other electrodynamical processes, such as electron scattering off atoms \cite{Wil2009}.

\medskip

In forthcoming articles, we will refine the presented results by including the cross terms which have been neglected in the calculation of the radial charge distribution \eqref{eq:Rho-IPM}.

\section{Acknowledgements}
This work was supported by the Research Council for Natural Sciences and Engineering of the Academy of Finland. 
All authors gladly thank Prof. Paul Indelicato, Prof. Nivaldo Lucio Speziali, and Prof. Carmelo Giacovazzo for discussions about possible applications of the presented results in structural crystallography. 
F.F. acknowledges support by Funda\c{c}\~ao de Amparo \`a Pesquisa do estado de Minas Gerais (FAPEMIG) and by Conselho Nacional de Desenvolvimento Cient\'ifico e Tecnol\'ogico (CNPq).
F.F. acknowledges hospitality of the N\'eel institute in Grenoble (France), where part of the work has been carried out.
P.A. acknowledges support by the German Research Foundation (DFG).
J.P.S. and P.A. acknowledge support by Funda\c{c}\~ao para a Ci\^encia e a Tecnologia (FCT) in Portugal, through the projects No. PEstOE/FIS/UI0303/2011 and No. PTDC/FIS/117606/2010.
%
%
%


\begin{thebibliography}{48}

\bibitem{Seth2013} Kamal K. Seth \etal, Phys. Rev. Lett. {\bf 110}, 022002 (2013).

\bibitem{Ahmed2012}	Z. Ahmed {\it et al.}, Phys. Rev. Lett. {\bf 108}, 102001 (2012).

\bibitem{Bernauer2011}	J. C. Bernauer {\it et al.}, Phys. Rev. Lett. {\bf 105}, 242001 (2010).

\bibitem{CoF2004} A. Courtoy, F. Fratini, S. Scopetta, and V. Vento, Phys. Rev. D {\bf 78}, 034002 (2008).

\bibitem{Chantler2000}	C. T. Chantler, J. Phys. Chem. Ref. Data {\bf 29}, 597 (2000).

\bibitem{Tar2002}	A. Tartari, A. Taibi, C. Bonifazzi, and C. Baraldi, Phys. Med. Biol. {\bf 47}, 163 (2002).

\bibitem{BaC1961}	B. W. Batterman, D. R. Chipman, and J. J. DeMarco, Phys. Rev. {\bf 122}, 68 (1961).

\bibitem{AlS2008}	A. Alatas \etal, Phys. Rev. B {\bf 77}, 064301 (2008).

\bibitem{Ces1992}	R. Cesareo \etal, Phys. Rep. {\bf 213}, 117 (1992).

\bibitem{Giac} C. Giacovazzo \etal, {\it Fundamentals of Crystallography} (Oxford Science Publications, 1992), chapter 2.

\bibitem{Wil2009}	D. B. Williams and C. B. Carter, {\it Transmission Electron Microscopy: 
						A textbook for material science} (Springer, 2009), chapters 2 and 3.

\bibitem{J.H.Hubbell:75} J. H. Hubbell \etal, J. Phys. Chem. Ref. Data {\bf 4}, 471 (1975).
\bibitem{J.H.Hubbell:79} J. H. Hubbell and I. \O verb\o, J. Phys. Chem. Ref. Data {\bf 8}, 69 (1979).

\bibitem{Crystallography:92} A. J. C. Wilson and V. Geist, {\it International Tables for Crystallography, Volume C: Mathematical, Physical and Chemical Tables} (Kluwer Academic Publishers, Dordrecht, 1992), chapter 6.1.

\bibitem{D.Belkic:83} D. Belkic, J. Phys. B {\bf 16}, 2773 (1983).

\bibitem{D.Belkic:84} D. Belkic, J. Phys. B {\bf 17}, 3629 (1984).

\bibitem{D.Belkic:89} D. Belkic and H. S. Taylor, Phys. Scr. {\bf 39}, 226 (1989).

\bibitem{D.Belkic:92} D. Belkic, Phys. Scr. {\bf 45}, 9 (1992).

\bibitem{R.T.Brown:70} R. T. Brown, Phys. Rev. A {\bf 1}, 1342 (1970).

\bibitem{R.S.Alassar:08} R. S. Alassar, H. A. Mavromatis, and S. A. Sofianos, Acta Appl. Math. {\bf 100}, 263 (2008).	

\bibitem{Schaupp:1983} D. Schaupp, M. Schumacher, F. Smend, P. Rullhusen, and J. H. Hubbell, J. Phys. Chem. Ref. Data, {\bf 12}, 467 (1983).

\bibitem{R.D.Cowan:81} R. D. Cowan, {\it The Theory of Atomic Structure and Spectra} (University of California Press, Berkeley, 1981).

\bibitem{W.Muhammad:2013} W. Muhammad, and S. H. Lee, PLoS ONE {\bf 8} e69608, doi:10.1371/journal.pone.0069608, (2013).

\bibitem{Bethe:1952} A. Bethe, Phys. Rev. {\bf 87}, 656 (1952).

\bibitem{Smend:1974} F. Smend and M. Schumacher, Nucl. Phys. A {\bf 223}, 423 (1974).





\bibitem{P.P.Kane:86} P. P. Kane, L. Kissel, R. H. Pratt, and A. C. Roy, Phys. Rep. {\bf 140}, 75 (1986).

\bibitem{B.H.Bransden:83} B. H. Bransden and C. J. Joachain, {\it Physics of atoms and molecules} (Longman 1983).

\bibitem{F.Fratini:2011} F. Fratini, \textit{One- and Two-photon decays in Atoms and Ions} (LAP, Saarbr\"ucken, Germany, 2011).

\bibitem{Andrey:2010}		A. Surzhykov, A. Volotka, F. Fratini, J.P. Santos, P. Indelicato, G. Plunien, Th. St\"ohlker, and S. Fritzsche, Phys. Rev. A {\bf 81}, 042510 (2010).

\bibitem{E.Clementi:1963} E. Clementi, and D. L. Raimondi, J. Chem. Phys. {\bf 38} 2686, (1963).

\bibitem{E.Clementi:1967}E. Clementi, D. L. Raimondi, and W. P. Reinhardt, J. Chem. Phys. {\bf 47} 1300, (1967).

\bibitem{webelement} http://www.webelements.com.

\bibitem{Slater:1930} J.C. Slater, Phys. Rev. {\bf 36}, 57 (1930).

\bibitem{A.Costescu:2011} A. Costescu, K. Karim, M. Moldovan, S. Spanulescu, and C. Stoica, J. Phys. B: At. Mol. Opt. Phys {\bf 44} 045204 (2011).

\bibitem{W.T.Howell:37} W. T. Howell, Philosophical Magazine (7) {\bf 24}, 1082 (1937).

\bibitem{L.Carlitz:61} L. Carlitz, Journal London Math. Soc. {\bf 36}, 399 (1961).

\bibitem{Abramowitz}	M. Abramowitz and I. A. Stegun, {\it Handbook of Mathematical Functions With Formulas, Graphs, and Mathematical Tables} (National Bureau of Standards, Applied Mathematics Series, 1964).

\bibitem{L.Kissel:80} L. Kissel, R. H. Pratt, and S. C. Roy, Phys. Rev. A {\bf 22}, 1970 (1980).

\bibitem{L.Kissel:00} L. Kissel, Radiat. Phys. Chem. {\bf 59}, 185 (2000).

\bibitem{Royetal}	S. C. Roy, L. Kissel, R. H. Pratt, Phys. Rev. A {\bf 27}, 285 (1983).

\bibitem{L.Safari:12A} L. Safari, P. Amaro, S. Fritzsche, J. P. Santos, and F. Fratini, Phys. Rev. A {\bf 85}, 043406 (2012).

\bibitem{L.Safari:12B} L. Safari, P. Amaro, S. Fritzsche, J. P. Santos, S. Tashenov, and F. Fratini, Phys. Rev. A {\bf 86}, 043405 (2012).

\bibitem{V.Florescu:90} V.~Florescu, M.~Marinescu and R.~H.~Pratt, Phys.\ Rev.\ A \textbf{42}, 3844 (1990).

\bibitem{Roy1999}			S.C. Roy, X-Ray Spectrom. {\bf 28}, 376 (1999).

\bibitem{RoyKissel:1999} S. C. Roy, L. Kissel, and R. H. Pratt, Radiat. Phys. Chem {\bf 56}, 3 (1999).

\bibitem{A.C.Mandal:02} A. C. Mandal, D. Mitra, M. Sarkar, and D. Bhattacharya, Phys. Rev. A {\bf 66}, 042705 (2002).

\bibitem{S.Kumar:09} S. Kumar, V. Sharama, J. S. Shahi, D. Mehta, and N. Singh, Eur. Phys. J. D {\bf 55}, 23 (2009).

\bibitem{E.Casnati:90} E. Casnati, C. Baraldi, and A. Tartari, Phys. Rev. A {\bf 42}, 2627 (1990).

\bibitem{P.Latha:12} P. Latha, K. K. Abdullah, M. P. Unnikrishnan, K. M. Varier, and B. R. S. Babu, Phys. Scr. {\bf 85}, 035303 (2012).

\bibitem{CDS}			{\it The Cambridge Structural Database (CDS)},
						http://www.ccdc.cam.ac.uk/products/csd/

\bibitem{ICDS}			{\it Inorganic Crystal Structure Database (ICSD)}, 
						http://www.fiz-karlsruhe.de/icsd\_content.html

\bibitem{PDB}			{\it Protein Data Bank}, 
						http://www.rcsb.org/pdb
\end{thebibliography}
\end{document}